\documentclass[conference]{IEEEtran}
\usepackage{cite}
\usepackage{enumerate}
\usepackage{amsmath,amssymb,amsfonts}
\usepackage{algorithmic}
\usepackage{graphicx}
\usepackage{textcomp}
\usepackage{xcolor}
\usepackage{comment}
\def\BibTeX{{\rm B\kern-.05em{\sc i\kern-.025em b}\kern-.08em
    T\kern-.1667em\lower.7ex\hbox{E}\kern-.125emX}}
\begin{document}

\title{
% VIFI UI: a Web-Based Graphical User Interface for Virtual Information Fabric Infrastructure\\
A Visual Analytics Framework for Distributed Data Analysis Systems
%= {\footnotesize \textsuperscript{*}Note: Sub-titles are not captured in Xplore and
% should not be used}

% \thanks{Identify applicable funding agency here. If none, delete this.}s
}
\def\plainauthor{\IEEEauthorblockN{Abdullah-Al-Raihan Nayeem $^{1}$, Mohammed Elshambakey$^{1,3}$, Todd Dobbs $^{1,2}$, Huikyo Lee$^{4}$, \\ Daniel Crichton$^{4}$, Yimin Zhu$^{5}$, Chanachok Chokwitthaya $^{5}$,  William J. Tolone$^{1}$, Isaac Cho$^{1,6}$}}
% \authoro Lee, Dan 
\author{\plainauthor\\ \IEEEauthorblockA{
 $^{1}$College of Computing and Informatics, University of North Carolina at Charlotte, Charlotte, United States \\ 
$^{2}$Computer Science, University of North Carolina at Greensboro, Greensboro, United States\\
$^{3}$City of Scientific Research and Technological Applications, Alexandria, Egypt\\
$^{4}$Jet Propulsion Laboratory, California Institute of Technology, Pasadena, United States\\
$^{5}$Construction Management, Louisiana State University, Rouge, United States\\
$^{6}$Computer Science, Utah State University, Logan, United States
}}
\maketitle

\begin{abstract}
%Data-driven decision-making has taken a momentous leap with the advancements in distributed data analysis. Growth in information technology infrastructures such as cheaper storage, better network speed, and larger bandwidth made the analysis across distributed and disparate datasets feasible. However, significant challenges remain for data owners and analysts to operate the data sites and conduct data analysis across the distributed servers. Imprecise server configuration, required to access multiple servers to run analysis, and develop interactive visualizations to explore the results bring undesirable complications for the analysts in their task. A web-based graphical user interface can play an essential role in managing access from the application layer and reduce user's direct interaction with the distributed servers. 

This paper proposes a visual analytics framework that addresses the complex user interactions required through a command-line interface to run analyses in distributed data analysis systems. The visual analytics framework facilitates the user to manage access to the distributed servers, incorporate data from the source, run data-driven analysis, monitor the progress, and explore the result using interactive visualizations. We provide a user interface embedded with generalized functionalities and access protocols and integrate it with a distributed analysis system. To demonstrate our proof of concept, we present two use cases from the earth science and Sustainable Human Building Ecosystem research domain.
% This document is a model and instructions for \LaTeX.
% This and the IEEEtran.cls file define the components of your paper [title, text, heads, etc.]. *CRITICAL: Do Not Use Symbols, Special Characters, Footnotes, 
% or Math in Paper Title or Abstract.
\end{abstract}


\section{Introduction}
To support decision-making in a data-driven society, research seeks to exploit the power of big data and the benefits of derived insights, scientific discoveries, and enhanced understanding. The advance and convergence of methods and technologies – including advances in machine learning and deep learning methods; increased storage capacities and reduced storage costs; higher network speeds and larger network bandwidth; more economical and powerful high-performance computing; and a growing prevalence of sensor networks and smart technologies – are essential enablers to enhanced sensemaking over big data. However, it is often the case that important insights and discoveries reside not within a single dataset, but instead are embedded within and across multiple and distributed datasets. Therefore, realizing the maximal potential for data-driven insights necessitates analyses and sensemaking that occur across these distributed, disparate datasets – analyses and sensemaking that, thereby, enable accurate and reliable revelation of latent, complex correlations, patterns, relationships, and such other knowledge that may not be revealed from a single dataset alone.

\begin{figure}
    \centering
    \includegraphics[width=\linewidth]{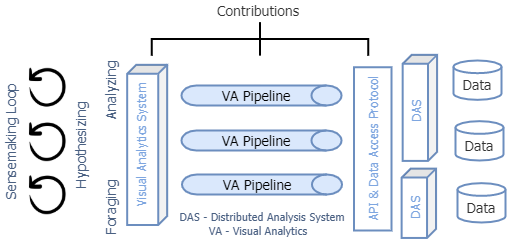}
    \caption{Visual analytic system pipelines for distributed analysis systems.}
    \label{fig:vifiui-introduction}
\end{figure}

There is an abundance of previous research (e.g., \cite{sandusky2016computational, cohn2012dataone, pordes2007open, medvedev2016sciserver, gesing2016using, foster2011globus, gugnani2016extending, talukder2017vifi}) spanning many disciplines that demonstrates the potential value and impact of enabling analyses and sensemaking across distributed, complex, and fragmented data. Yet, significant challenges remain. In particular, to support sensemaking across such data, new visual analytic interfaces are needed, new pipelines for optimized distributed data interaction and visualization are required, and new data access protocols and application programmer interfaces (APIs) must be developed. We highlight these challenges in Figure \ref{fig:vifiui-introduction}.

In support of sensemaking, users require a visual analytic interface that seamlessly supports data discovery, exploration, and analyses. In other words, the visual analytic interface should support the full extent of the sensemaking loop \cite{pirolli2005sensemaking, sacha2014knowledge} from foraging to hypothesizing to analyzing. Current solutions, however, often emphasize specific aspects of sensemaking – for example, data exploration or data analyses – and fail to support the full analytical lifecycle adequately. In addition, it is infeasible to access large and remote datasets using traditional pipelines for data transformation, conversion, and presentation. Such pipelines are commonly preceded by massive data downloads, which are infeasible or impractical for many remote datasets. Thus, new pipelines are required, pipelines that are not predicated on massive data downloads. Finally, to generalize visual analytic interface for distributed fragmented data, new APIs and data access protocols are necessary. In particular, these APIs and protocols must account for the full analytical lifecycle and must not be predicated on massive, upfront data downloads.

In this paper, we present an interactive visual analytics framework (VAF) for the distributed data analysis systems (DAS). VAF enables analyses over distributed, fragmented data without the movement of massive data. Significant advancements in distributed data analysis over the past decade \cite{talukder2017vifi, medvedev2016sciserver, 6866038, 7840984, Shahand2014} make our proposed framework a feasible candidate to accelerate the analysis tasks of researchers and analysts. To demonstrate our framework, we leveraged the Virtual Information-Fabric Infrastructure (VIFI) \cite{talukder2017vifi, elshambakey2017towards, chokwitthaya2019combining, karaguzel2019open, zhang2020development, j2019application, bhattacharjee2019multi}, which is a computational infrastructure that enables analyses across distributed, fragmented data without the movement of massive data. Within VIFI, analyses migrate to the distributed data and only derived data – e.g., result sets – migrate from the data hosts. 
Our contributions of this paper are:

\begin{itemize}
    \item We define a VAF for distributed, fragmented data as well as design goals and associated implementation tasks.

    \item We present a generalized pipeline for data transformation, conversion, and presentation – one that is not predicated on massive, upfront data downloads.

\item We provide a demonstration version of the visual analytic user interface (UI) to support distributed analysis. 

    \item We present generalized APIs and data access protocols to enable proper integration with infrastructures that enable analytics over distributed fragmented data.

    \item We demonstrate VAF with two analytic systems (i.e., VIFI and a simple file-based systems) and illustrate its benefits using two uses cases from earth science and Sustainable Human Building Ecosystem (SHBE) research domains. 

\end{itemize}

\section{Related Work}
Current data-driven applications often require the identification and mitigation of relevant data from multiple locations to a common storage location, prior to performing analysis. To overcome what is often a difficult, time-consuming, and laborious task, some alternate solutions have been proposed for data sharing using high-speed networks and cloud-based hosting, while other alternative solutions focus on providing shared computing resources. DataONE \cite{sandusky2016computational, cohn2012dataone} is a project focused on providing easier access, search and discovery to earth and environmental science data repositories. The Open Science Grid \cite{pordes2007open, Sfiligoi:2009:PWG:1579192.1579413} enables scientific research by
providing distributed computing resources. SciServer
\cite{medvedev2016sciserver,doi:10.1177/0002716217745816} is a cyber-infrastructure system that provides a suite of tools and services (including storage, access, query, and processing) for big data analyses from various disciplines leveraging data with different format and structure. While SciServer collects all data at a common storage location, it attempts to minimize data movement by collecting data at the location that contains the majority of the required data. SciServer also migrates the analyses by sending Jupyter Notebook \cite{jupyter_notebook_shambakey} to the common storage location.

Other data-driven applications aim to develop research infrastructures that integrate storage, high-performance computing, and analytic tools (e.g., XSEDE \cite{6866038,doi:10.1002/cpe.1098}, NeCTAR \cite{7881699}, PRACE \cite{prace_shambakey}, and EGI \cite{egi_shambakey}). The applications allow end-users to share distributed computing resources and data repositories. The solutions may be used by Science Gateways (SGs) \cite{Gottdank2014,gesing2016using,Piontek2016,Farkas2016,GESING201997} to provide (web) portals and UIs that enable scientists (e.g., chemists, biologists) to access, build and execute analytic workflows. SGs relieve scientists of the burden and needed expertise to setup and maintain the underlying distributed cyber-infrastructure. SG services can be shared and reused by different end-users. SGs can be classified into SG framework like WS-PGRADE/gUSE \cite{Gottdank2014}, and  SG instances like the computational neuroscience gateway \cite{Shahand2014}. SG frameworks are generic SGs that provide low-level services for scientists from different domains. While SG frameworks provide high-level abstractions for computing specialists, SG frameworks require additional learning from the scientists to leverage the full potential of the frameworks. SG instances provide high-level services for scientists in a specific domain. Thus, SG instances simplify scientific operations for end-users, but limit flexibility when more functionalities are needed from the SG instance. Some of the SG features and
services (e.g., security, data and workflow management) depend on the underlying technology. Thus, it becomes challenging to port a SG from one infrastructure to another
\cite{7217932,Arshad2016}. Gugnani et al. \cite{Gugnani2016} suggests a generic approach to integrate
infrastructure aware workflows, (e.g., WS-PGRAD/gUSE \cite{Gottdank2014}) with bigdata parallel processing tools (e.g., Hadoop). This work \cite{Gugnani2016} uses the CloudBroker platform \cite{7217927} to provide required cloud-based computational resources.

SGs can be accessed through different middleware like Airavata \cite{6882068}, Agave \cite{agave_shambakey}, and Globus \cite{5755602,Chard:2016:GRE:2949550.2949554,Allen:2012:SSD:2076450.2076468}. Airavata \cite{6882068} allows users to manage applications and workflows on the provided resources (e.g., clouds, cluster, grids) through component abstraction of major tasks. The system components are indirectly accessed through component APIs. Agave \cite{agave_shambakey} provides web-access, through Representational State Transfer (RESTful) APIs \cite{fielding2000architectural}, to given resources (e.g., HPC, cloud) to run analyses and  and to manage data.
Globus \cite{5755602,Chard:2016:GRE:2949550.2949554,Allen:2012:SSD:2076450.2076468} is software-as-a-service designed to make it easier to discover, replicate, and access big data resources at different locations. Globus is used to deliver scalable research data management services in a secure manner to a variety of stakeholders. Some Globus features, like data publication and managed endpoints, include licensing fees.

In contrast to existing solutions, our VAF aims to support ``truly distributed analytics'' where analytics are executed at data sites without the massive movement of data. Our framework avoids huge data transfer times while complying with owner-defined authentication and authorization policies for data access. Our framework does not add new infrastructure for additional data and/or computational operations; rather, it aims to integrate with existing data site infrastructure. The framework utilizes containerization technology (e.g., Docker \cite{7093032,7036275,7307636,7377291,7506647,Boettiger:2015:IDR:2723872.2723882,Dikaleh:2016:HBP:3049877.3049914,Merkel:2014:DLL:2600239.2600241,Miell:2016:DP:3052489}), rather than tools like Jupyter notebooks \cite{jupyter_notebook_shambakey}, to migrate analyses. This provides more flexibility over the analytics tools and analytic environments that can be used by the scientists in conducting data-driven inquiries (i.e., analyses are not limited to the tools provided by Jupyter). In addition, unlike some related work, our framework depends entirely on open source technology. For example, our pipeline uses only open-source components (e.g., Apache NiFi \cite{nifi_shambakey} and Docker Swarm \cite{docker_swarm_shambakey}) with free access to all features. Thus, users can develop, reuse, and customize our framework for their needs. %

\section {System Design}
    \begin{figure*}[t]
        \centering
        \includegraphics[width=\linewidth]{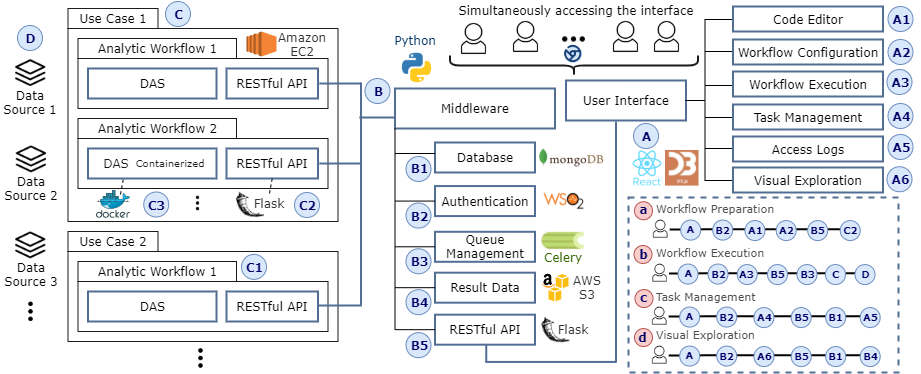}
        \caption{Proposed VAF pipeline for DAS: A) User Interface, B) Middleware, C) DAS site, and D) Data site are the main modules in our framework. a) Workflow Preparation, b) Workflow Execution, c) Task Management, and d) Visual Exploration show the flow of interactions within the VAF components.}
        \label{fig:system-pipeline}
    \end{figure*}

This paper proposes an interactive VAF to simplify user interactions and enhance the user experience with a DAS. To design a pipeline for the VAF, we reviewed numerous distributed analytic systems (e.g., \cite{talukder2017vifi, medvedev2016sciserver, potter2009visualization, nifi_elshambakey}) to identify the key user interactions required to operate these systems. We discovered that many system utilized command line interfaces. Nonetheless, we extracted the following fundamental interactions: managing access to distributed servers, preparing analytic scripts and runtime environments, importing data from remote sources, executing analyses, monitoring the execution progress, and inspecting and exploring the analytical results. DAS commonly maintain data site to data site communication using cloud infrastructures to run analyses \cite{talukder2017vifi, potter2009visualization, elshambakey2017towards}. To operate a DAS from a command line interface requires access for a user to multiple remote servers. Access control for such interaction with the data sites and DAS sites can be complex for the data owners. Consequently, the entire procedure to run a data analysis can be similarly challenging for the data analysts and the end users. Moreover, to explore the results, users from different domain areas were required to pull the resulted data from the server. Rather than using command line interfaces, DAS often provide a visualization toolkit \cite{potter2009visualization, 6407220e803d42bfb42f532936a673d6}. However, users are responsible for generating the exploratory visualizations or necessary artifacts to measure the performance of the analysis \cite{SHAKHOVSKA2019561}. Given all of these need interactions and associate limitations of current solutions, we identified associate design requirements and implementation tasks and mitigate current complexities for user-DAS interaction.

\subsection{Design Requirements}
We propose an interactive VAF to provide more seamless user interaction with distributed analysis systems. Related work reveals the following design requirements for our VAF:

\begin{enumerate}[DR1]
    \item \textbf{To mediate user interaction with distributed servers.} The framework should provide sufficient features to allow users to execute analyses in DAS without requiring direct user access to the distributed servers and data hosts.
    
    \item \textbf{To provide a unified model for authentication and access control for distributed servers.} The framework should provide proper access to data and analytic workflows according to data site policies. The framework should integrate with existing authentication and authorization mechanism to the computing servers and various data sites. 
    
    \item \textbf{To enable the exploration of data and resulting analyses using interactive visualizations} The framework should utilize interactive visualizations to support the sensemaking loop (i.e., foraging, hypothesizing, and analyzing) while not requiring massive data downloads as a means to enable accurate and reliable revelation of latent, complex correlations, patterns, relationships, and such other knowledge.
\end{enumerate}

\subsection{Implementation Requirements}
To address the above design requirements, we identify the following implementation requirements for our framework:
\begin{enumerate}[R1]
    
    \item \textbf{To provide an interface to manage analytical scripts and Portable Analytic Containers (PACs).} Framework users must be able to access, specify, and manage analytical scripts that are stored in an external repository – e.g., at a DAS data host. As such, the framework should offer an end-to-end synchronization with the available analytic scripts and PACs in DAS \textbf{(DR1)}.
    
    % \item Synchronize with the PAC registry to keep the scripts up-to-date
    \item \textbf{To enable user efforts to configure analytical scripts and workflows.} To conduct analysis across distributed, fragment data, coordinated execution of analytical scripts is often required (hypothesizing). Workflows often contain a set of configurations that points the dataset, analysis scripts, required access credentials, etc. The framework should provide affordances for users to modify analytical workflow configurations \textbf{(DR1)}.
    
    \item \textbf{To support user-initiated execution of analytical workflows in DAS.} After enabling the preparation analysis scripts and configuring an analytical workflow, the framework should allow the user to initiate workflow execution. In addition the framework should minimize the need for the user to authenticate directly to each data host (e.g., mediate authentication via single sign-on) \textbf{(DR1, DR2)}.
    
    \item \textbf{To mediate and comply with data host authentication requirements and authorization policies for datasets, analysis scripts, and workflows.} The framework should manage compliance with authentication requirements and authorization policies for end users. Users should be able to view, modify, and execute analysis scripts and workflows on permitted datasets according to data host authorization policies \textbf{(DR2)}.
    
    \item \textbf{To maintain user awareness of workflow execution status.} Workflows often require significant time to queue and execute. The framework should maintain user awareness of workflow execution status so that users may accurately track their progression in the DAS \textbf{(DR1)}.
    
    \item \textbf{To provide access to the runtime and error logs.} Runtime logs are useful for the users to understand DAS performance and anticipate expected runtimes of analytical workflows. Similarly, error logs are helpful to trace script and workflow execution, particularly in exceptional circumstances. The framework should effectively present runtime and error logs to users \textbf{(DR2, DR3)}.
    
    \item \textbf{To provide an interactive visual analytic interface to support data discovery and explore analytical results.} The framework should provide users interactive visualizations to discover data (foraging) and explore workflow results (analyzing). The visualizations may be general-purpose or analysis-specific. Thus, the framework should be extensible to accommodate analysis-specific visualizations \textbf{(DR3)}. 
\end{enumerate}

To satisfy the design and implementation requirements for the proposed VAF for DAS, we developed: interactive, web-based, visual analytic interfaces; a visual analytic pipeline; and, an API / data access protocol.

\section{Middleware}
The middleware for the VAF is one of two major components of the visual analytics pipeline as well as the implementer of the data access API and protocol (Figure \ref{fig:system-pipeline}B). It orchestrates use case \textbf{(R1)}, workflow/task \textbf{(R1-3, R5-6)}, and script \textbf{(DR1, DR2)} management as well as authentication \textbf{(R4)} and authorization integration \textbf{(R4)} with DAS. The primary component is the Task Manager that mediates communications between the VAF and DAS.  
% Figure \ref{fig:middleware} summarizes the primary functions of the VAF middleware. Each is discussed in greater detail in the following. 

% \begin{figure}
%     \centering
%     \includegraphics[width=\linewidth]{figures/middleware.png}
%     \caption{DAS Middleware Architecture}
%     \label{fig:middleware}
% \end{figure}

\textbf{Use Case Management:} Use case management provides methods for creating, modifying, and projecting use cases. A use case organizes a collection of analytical workflows and results. The Task Manager creates a unique key for each new use case. The user, then, specifies and name and one or more workflows (Figure \ref{fig:system-pipeline}B1). Results from executed workflows are also collected in a use case. As such, use cases provide a means to organize analyses.

\textbf{Workflow/Task Management:} Workflow/task management provides methods for the creation, mutation, execution, and projection of workflow specifications and execution instances. Executing workflows are called tasks. Each task is attributed with script identifiers, user identifier, use case, and workflow. When a user submits a request to execute a workflow, a new task is created and scheduled for execution via the DAS \textbf{(R2, R3)}. The Task Manager collects the required information and relates the information to a unique identifier corresponding to its workflow and use case, respectively. The task along with its related scripts are, then, sent to the DAS for execution (Figure \ref{fig:system-pipeline}a). Task information, including execution steps and status updates, is captured in the runtime and error logs \textbf{(R5, R6)}. Once a task completes, the Task Manager retrieves the analytical results from the DAS.

\textbf{Script Management:} Script management provides methods for the creation, mutation, and projection of scripts. Scripts and their related configurations are associated with each workflow/task. The script identifier is used during the task creation process to ensure all relevant analyses are properly identified and subjected to the DAS for execution \textbf{(R1)}. The 
alytics interface leverages use case, workflow/task, and script management collectively in the Task Manager to support hypothesizing activities as part of the sensemaking loop.

\textbf{Results Management:} Result management provides methods for the projection of analytical results (e.g., task results). Results for each workflow are associated with a task identifier. When the execution of a workflow completes, the DAS signals the completion status to the Task Manager \textbf{(R5)}. The Task Manager, then, retrieves results from the DAS so that these may be projected to the user via the visual analytics interface \textbf{(R7)}.

\textbf{Authentication:} To meet the VAF authentication requirements, the middleware uses InCommon \cite{incommon} and WSO2 \cite{wso2_is} for identity management. The VAF, leveraging these services, implements key-based authentication to enable trusted communication between VAF and DAS components \textbf{(R4)}.

InCommon is a federated identity management service provided to education and research institutions using the Shibboleth single sign-on architecture. Given the large number of participating institutions and simplicity of setup, VAF integrates with InCommon-based authentication services \cite{incommon}.

For users whose institution is not a member of the InCommon federation, the WSO2 Identity Server (IS) is utilized for authentication. The WSO2 IS integrates with any IAM-compliant architecture. For users with no IAM-compliant architecture, WSO IS provides a built-in IAM architecture. While WSO2 integrates with Shibboleth SSO, and thus may be integrated with InCommon, the current VAF implementation leverages InCommon outside of WSO2 IS to simplify configuration \cite{wso2_is}. 
For VAF configurations that leverage WSO2 IS, the middleware authorization service uses the WSO2 API to handle user authorization requests. In such implementations, user authorizations are configured using the WSO2 IS Administration application.

Within VAF, key-based authentication enables trusted communication among VAF components and between VAF and DAS. For WSO2 implementations of VAF, key-based authentication also is enabled between middleware services and WSO2 services. Key-based authentication leverages Hypertext Transfer Protocol Secure (HTTPS) and requires valid certificates for communication between endpoints. 

\textbf{Authorization:} The middleware provides two options for authorization support, either a proprietary solution or a WSO2 implementation. %\subsubsection{Proprietary}
For VAF configurations that do leverage WSO2, a proprietary authorization solution is provided  via a middleware authorization service. To set up user authorizations using the service requires manual database updates. 

We defined three user roles for VAF:  a data owner, workflow designer, and data analyst \textbf{(R4)}. A data owner manages the user's access control in DAS data sites. A Workflow designer setup an initial configuration and orchestration path for a new workflow. 
% The designers ought to obtain the technical specification of VAF. 
While a data analyst is authorized tweak certain configurations according to need, the designer's role is to use the workflows to conduct analyses.

\begin{figure*}[t]
    \centering
    \includegraphics[width=\linewidth]{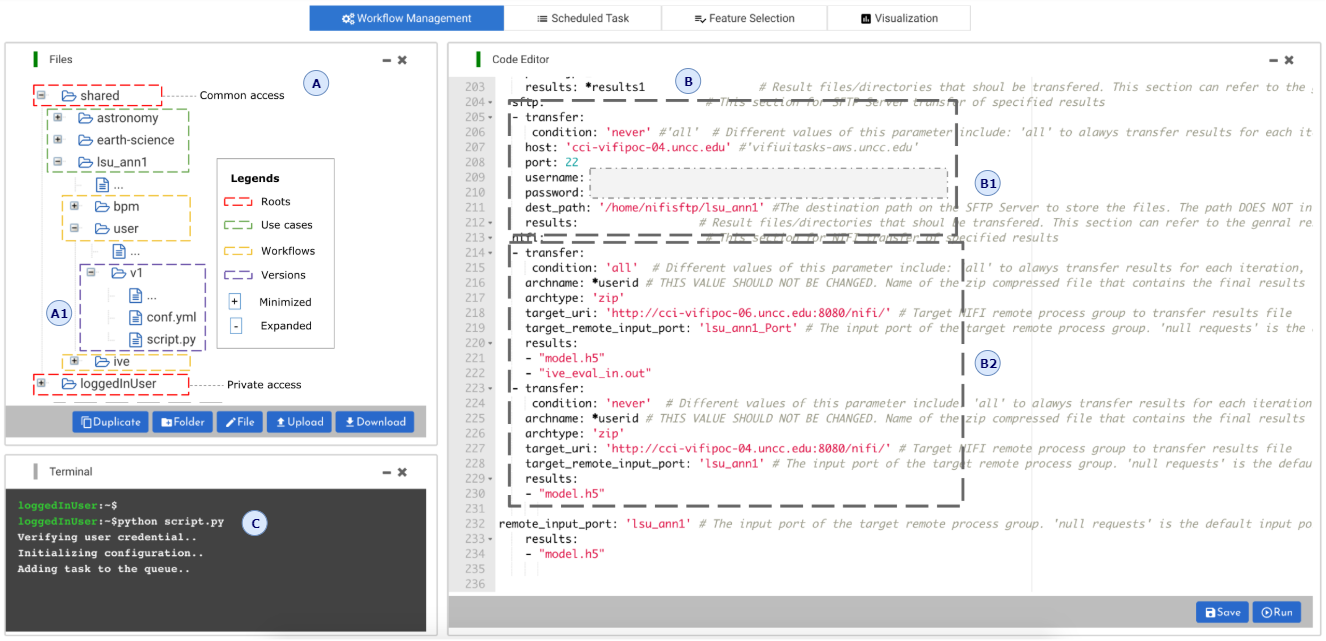}
    \caption{Analytic workflow management through our visual analytics systems. Workflow management view provides A) File Browser - to configure the workflow, B) Code Editor - to prepare and run the analytic workflows, and C) Terminal - to stream raw outputs.}
    \label{fig:workflow-management}
\end{figure*}

\section{Visual Analytics Interface}
The visual analytics interface is the second major component of the visual analytic pipeline. It provides coordinated views \cite{andrienko2007coordinated} to support user actions for workflow execution and result exploration in DAS. To satisfy the design requirements, the UI introduces three main panels: workflow management, task management, and result exploration. These panels assist users in three different phases of sensemaking: 1) data exploration (foraging); 2) analytical workflow and script development and execution (hypothesizing); and, 3), exploring and analyzing workflow results (analyzing). In the following sections, we illustrate support for each phase by presenting VAF support for workflow/script management (hypothesizing), task management (hypothesizing), and interactive visual exploration (foraging and analyzing).

\subsection{Workflow/Script Management}
The workflow/script management view consists of a File Browser, a Code Editor, and a Terminal View (Figure \ref{fig:workflow-management}A,B and C). In the File Browser, the available use cases and workflows are listed according to user access privileges to the PAC repository \textbf{(R1, R4)}. By default, the view provides access to two types of directories: shared directories and user directories. The shared directories contain all use cases and workflows that are shared with other users. The user directories contain the use cases and workflows (created or cloned) that are private to the user. The File Browser is synchronized with the middleware's Task Manager component via RESTful API. The workflow/script management view presents only those use cases and workflows that are configured in the DAS and flagged as enabled in the Task Manager. We require hierarchical presentation of PACs in the associated DAS as shown in Figure \ref{fig:workflow-management}A. The hierarchy is set in a manner that always gives an ordered path (\texttt{/[ root-directory ]/[ use-case ]/[ workflow ]/[ w-version ]/}) when users select a workflow to execute. For example, if the user decides to execute the version 1 of the user workflow shown in Figure \ref{fig:workflow-management}A, the conceptual path to the script directory would be \texttt{/shared/lsu\_ann1/user/v1/}. The hierarchical abstract organization is adopted for its familiarity and ease of use. Moreover, it provides an encoding that facilitates interface middleware communications.

% \begin{figure}
%     % \centering
%     \includegraphics[width=\linewidth]{figures/file-structure.png}
%     \caption{Hierarchical file structure for the use cases and workflows in the UI. Workflow options are rendered based on user's access on the repository.}
%     \label{fig:file-structure}
% \end{figure}
We added operations to the File Browser (Figure \ref{fig:workflow-management}A) to create, duplicate, or modify the workflows \textbf{(R1)}. To keep the integrity of the file structure, each operation is implemented with a set of constraints (Figure \ref{fig:system-pipeline}a). The "Duplicate" operation allows the user to clone a selected workflow. It also allows users to clone scripts. For example, in Figure \ref{fig:workflow-management}A1, \texttt{ive2.py} is duplicated (or cloned) from \texttt{ive1.py}. However, this operation does not allow users to clone use cases or the root directory. Similarly, "Add folder" only allows users to create new version folders under a selected workflow, rather than creating a folder at an arbitrary location in the hierarchy. The "Upload" and "Download" actions allow the user to migrate analysis to and from the local machine and the DAS. 

In the Code Editor (Figure \ref{fig:workflow-management}B), the user can modify the workflow, and create and modify scripts according to their hypotheses for the corresponding use case. By selecting a script, users are allowed to modify and execute the script within the Code Editor for testing purposes (Figure \ref{fig:workflow-management}B). The File Browser also provides access to workflow configurations, which users can select to modify in the Code Editor \textbf{(R2)}. In the File Browser (Figure \ref{fig:workflow-management}A), the scripts and workflow configurations are validated prior to execution to assess whether modifications are permitted. The \texttt{conf.yml} file associated with each workflow version contains the workflow specification and identifies the appropriate DAS for execution. This file includes, among other things, the DAS credentials (Figure \ref{fig:workflow-management}B1), dataset identifiers, and the location where workflow results are to be transferred after task execution completes (Figure \ref{fig:workflow-management}B2). The Terminal (Figure \ref{fig:workflow-management}C) reflects the output from an associated command line interface to the DAS (when such an interface exists). It also shows log files and the output of test script executions.

To execute a workflow in a DAS, the user selects the \texttt{conf.yml} file for the workflow in the File Browser (shown in Figure \ref{fig:workflow-management}A1) and clicks the "Run" button located bottom right in the Code Editor (Figure \ref{fig:workflow-management}B) \textbf{(R3)}. The interface, then, passes the command to the middleware and switches to the Task Management view once execution is launched in the DAS (Figure \ref{fig:system-pipeline}b).

\begin{figure*}[t]
    \centering
    \includegraphics[width=\linewidth]{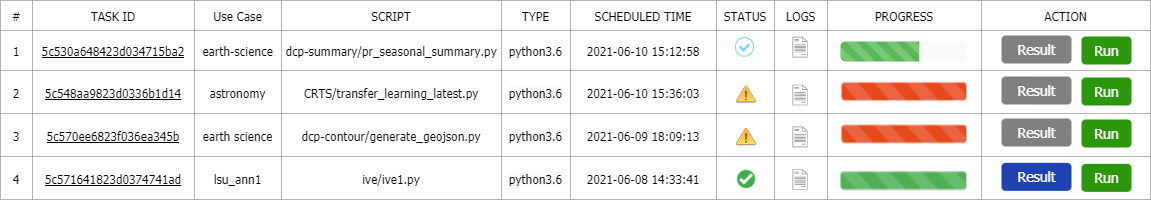}
    \caption{The UI for submitted analysis task management in our visual analytics system. Task management interface allows the user to interact with the scheduled tasks for inspecting logs, tracking progress, visual exploration of the results or re-running the workflow.}
    \label{fig:task-management}
\end{figure*}

\subsection{Task Management}
The task management view contains a Scheduled Task panel that lists the workflows (i.e., tasks) that are currently executing for the given user as shown in Figure \ref{fig:task-management}. The Scheduled Task panel provides graphical indicators of task progression. A unique task ID is generated for each workflow execution \textbf{(R5)}. While executing the workflow, the task identifier is linked to all runtime data, including the runtime environment, script directory, logs, results files, etc \textbf{(R6)}. 

A task may take anywhere from fractions of a second to hours or days to execute depending on the size of the data, the complexity of the analysis, the computational resources available, and the shared demand for the data and computing resources. While a task is executing, the user can interact with any tasks to inspect execution logs or view the results of completed tasks (Figure \ref{fig:system-pipeline}c). The execution logs accessible from the Task Manager are not the output logs from the given script. Rather, these logs, retrieved from the middleware, capture workflow progression checkpoints for a given task, such as: a) queued – execution request sent to middleware; b) queuing - middleware retrieving relevant scripts, preparing for task execution, generating the unique  task identifier, etc.; c) created – the workflow execution request is validated the request and the task is properly created; d) sending - transferring the task to the appropriate DAS; e) sent - the task is successfully sent to the DAS and awaiting execution, and f) complete - the DAS completed the task execution and results are returned to the middleware for user access.

The progress bar aligned with each task in the table (Figure \ref{fig:task-management}) depicts an estimation of overall execution progression. The Scheduled Task panel provides users with several operations that may be applied to a given task, including: a) “Cancel” – this operation allows a user to cancel task execution by the DAS, b) “Rerun” – this operation allows a user to rerun a task, possibly with updated parameters, after first canceling the current execution; and, c) “Result” – this operation, available after task completion, takes a user to interactive visual interfaces to explore the data that result from task execution.

% FUTURE WORK - With the interface, the user can provide information to prioritize the execution of a workflow that essentially manages the queue.
% FUTURE WORK - allows user to explore the result on the go..

\subsection{Visual Exploration}
\begin{figure}[t]
    \centering
    \includegraphics[width=\columnwidth]{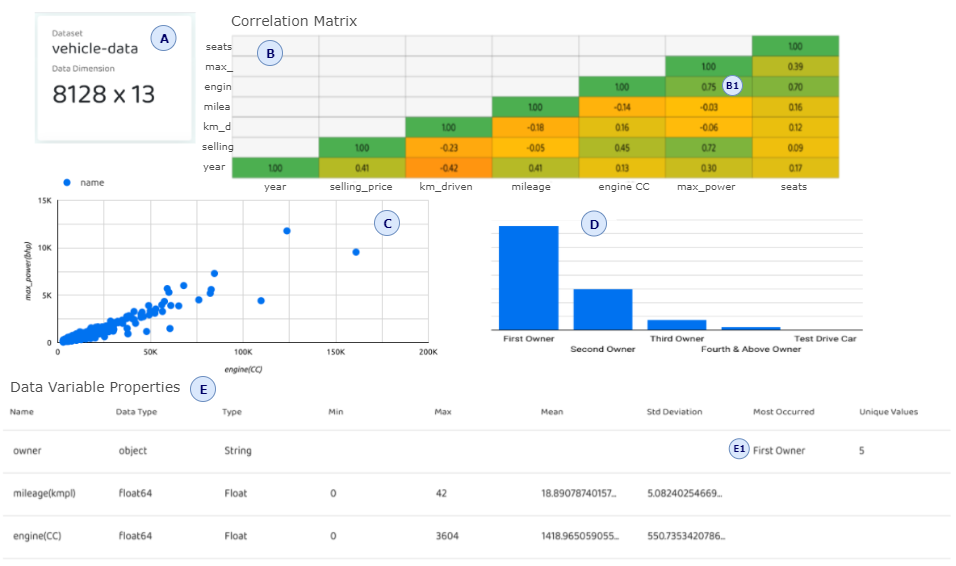}
    \caption{The variable exploration panel on the UI for analysis results. A) Metadata, B) Triangle matrix - variable correlation, and C) Data variable properties familiarize the user with the resulted.}
    \label{fig:result-exploration}
\end{figure}

The interactive, visual exploration views provide a threefold means to explore both data/datasets (foraging) and task results (analyzing). In this section, without loss of generality, we focus our presentation on results exploration (Figure \ref{fig:system-pipeline}d). The interactive, visual exploration views include two principal panels: the variable exploration panel and visual exploration panel \textbf{(R7)}.

The variable exploration panel provides a view that allows users to explore the properties of resulted data. Figure \ref{fig:result-exploration} shows a sample illustration of the variable exploration panel using this data \cite{KrauseStarkDengFei-Fei_3DRR2013}. The data variable exploration panel initially provides the data dimension (Figure \ref{fig:result-exploration}A), a triangle matrix (Figure \ref{fig:result-exploration}B) and a data table containing the variable properties (Figure \ref{fig:result-exploration}E). We implemented this panel recognizing that users may not always be familiar with the data variables. This panel provides the data type for each variable in the data. In addition, for numeric data variables, the table provides some statistical data (e.g., range, mean, and standard deviation), though this may not always be relevant or useful. For categorical data, the panel provides count and frequency information. For example, hovering over categorical data presents a bar chart providing the frequency distribution of the categorical data. Additionally, the matrix (Figure \ref{fig:result-exploration}B) provides the correlation among data variables, which may help users during analyses. The matrix cells are color coded and denotes the correlation -1 to +1 using a red-yellow-green color scheme. The user can explore the correlation between two variables by hovering the mouse over the corresponding cell in the triangle matrix. The scatter plot and bar chart (Figure \ref{fig:result-exploration}C, D) based on the respective interactions with variable properties (Figure \ref{fig:result-exploration}B1, E1) allow users to identify and explore patterns or outliers in the data.

% \begin{figure}
%     \centering
%     \includegraphics[width=\linewidth]{figures/feature-transformation.png}
%     \caption{VIFI resulted variable transformation \& visualization recommendation settings}
%     \label{fig:feature-transformation}
% \end{figure}

The data transformation capabilities include scaling the data variables, applying statistical summary or formula to transform data variables, and injecting domain knowledge to nudge the exploration panel in identifying relevant visualizations. 
% For each variable, the user can scale the values within a range, group the data to calculate summation and mean for the numeric variables and total count for categorical variables. 
% Moreover, the user can insert rules to filter the resulting data samples and mark variables as important. 
Additionally, the UI allows the user to input thresholds such as good, moderate, and poor correlations, standard deviations, and minimum and maximum factors for the unique values that are perceived as the user's domain knowledge. 
% The visualizations are recommended based on the variable's significance and the user injected domain knowledge. 
The user can save the action items as a transformation profile to apply in the future resulting data from the workflow.

The interactive, visual exploration panel provides a view that recommends visualizations methods to users based on the data type and format. Users may also independently select relevant visualizations from the palette of available visualization methods. This palette is also extensible to allow users to add highly tailored visualizations for specialized data or analysis tasks. This latter feature is provided in recognition of anticipated unconventional visualization requirements for different varying use cases \textbf{(R7)}. To support interactive, visual exploration, we modularized the exploration panel based on the use case. As such, the visual exploration panel for each use case inherits the common visualizations and includes (optional) custom visualizations. For example, to support the sensemaking in one use case (discussed in Section ~\ref{sec:use case 1: LSU_ANN}), we implemented the interactive custom scatter plot shown in Figure \ref{fig:lsu-use-case-re}. The inherited visualization library  includes line charts, standard scatter plots, parallel-coordinates, %pie charts, 
box plots, heat maps, geospatial maps, and tabular data presentations.

\section{Distributed Analysis System: VIFI}
To evaluate VAF, we integrate VAF with two DAS: a simple file-based DAS and the Virtual Information Fabric Infrastructure (VIFI) DAS. In this section, we describe the latter DAS which serves as the foundation of most of our VAF evaluation activities.

VIFI \cite{talukder2017vifi, elshambakey2017towards, chokwitthaya2019combining, bhattacharjee2019multi} is a DAS that enables analyses across distributed, fragmented data without the movement of massive data. Within VIFI, analyses migrate to the distributed data and only derived data – e.g., result sets – migrate from the data hosts. VIFI supports research and analysis in multiple domains including astronomy \cite{bhattacharjee2019multi}, earth science \cite{talukder2017vifi}, and sustainable human-building ecosystems (SHBE) \cite{chokwitthaya2019combining}. The current implementation of VIFI consists of the following components: Portable Analytic Containers, Registry Services, Orchestrator, User Node, and Data Sites.
%, Metadata Server, Crawler, and Watchdog. 
Each is described briefly in the following.

% \begin{itemize}
    % \item 

    \textbf{Portable Analytic Containers (PACs):} A PAC is a lightweight virtual machine, called a container, that hosts software, libraries, %and any other tools
    and operating system needed by end users to analyze data. A PAC can receive and execute analysis programs (e.g., scripts) if the required programs are not already contained in the PAC. Leveraging container technology (e.g., Docker~\cite{{7093032,7036275,7307636,7377291,7506647,Boettiger:2015:IDR:2723872.2723882,Dikaleh:2016:HBP:3049877.3049914,Merkel:2014:DLL:2600239.2600241,Miell:2016:DP:3052489}}). A PAC is portable to migrate and execute on heterogeneous host platforms. A PAC facilitates reusability by hosting and utilizing different analytical libraries and programs pulled from shared repositories (e.g., Docker hub~\cite{dockerHub_shambakey}). Container technology enables the movement of analytics rather than the movement of data; thus, alleviating problems related to the transfer of big data. PACs offer a number of affordances for distributed analytics: i) they can be easily transmitted over the network due to their limited size; and ii) they simplify analytics development for inexperienced users. The VIFI infrastructure is scalable as it enables the integration of various VIFI nodes at different sites. The ability for VIFI workflows to access fixed sites allows VIFI to cooperate with non-open-source resources, assuming that a VIFI user has the proper credentials. Currently, VIFI researchers are extending VIFI to use Singularity~\cite{10.1371/journal.pone.0177459,DBLP:journals/corr/abs-1709-10140,Le:2017:PAA:3093338.3106737} to run on High Performance Computing (HPC) clusters at different sites.
 
    % \item 
    \textbf{Registry Services:} Distinctive PACs are stored, searched, utilized and shared through Registry Services. Currently, VIFI uses Docker hub~\cite{dockerHub_shambakey} to implement the Registry Services. We expect future VIFI versions to incorporate additional services to advance download and transfer times of PACs.
    
    % \item 
    \textbf{Orchestrator:} The Orchestrator automatically coordinates workflow (i.e., task) execution across multiple VIFI sites (i.e., distributed datasets).  Each analysis step in a workflow is implemented by a script running in a PAC at a data site. Although initial VIFI implementations used NiFi~\cite{nifi_elshambakey,nifi_shambakey} as its orchestrator, current implementation use RESTful APIs to improve orchestrator customizability.	
    
    % \item 
    \textbf{User Node:} The user node is the means by which users interact with the VIFI framework. The user node provides a user interface, communication, and basic computation capacities.

    % \item 
    \textbf{Data Site:} Data Sites are locations in the VIFI infrastructure  where distributed, fragmented data reside. Each VIFI Data Site interacts with the Orchestrator (i.e., NIFI and/or RESTful APIs) and runs PACs (e.g., by Docker Swarm~\cite{7753148}). VIFI uses Docker Swarm to execute parallel analytics. Each Data Site runs a VIFI server supported by a configuration file that configures hosted data sets and log files at this site.
\begin{comment}    
    % \item 
    \textbf{Metadata Server:} The Metadata Server stores and lists gathered metadata about datasets at each Data Site. These metadata are used to support data discovery (foraging).

    % \item 
    \textbf{Crawler:} The Crawler is used by the Metadata Server to automatically collect metadata at each Data Site.

    % \item 
    \textbf{Watchdog:} The Watchdog updates the Metadata Server when modifications to the metadata are detected at any of the associate Data Sites.
\end{comment}

VIFI workflows are either launched from the command line interface of the VIFI server running at each Data Site or via the User Node. The VAF reported in this paper functions as the VIFI User Node for the use case evaluations reported in the following section that used VIFI.

\section{Use cases}
To evaluate the affordances of our visual analytics framework, we implemented the framework leveraging the VIFI DAS. As part of our evaluation, we present two use cases: one from the earth sciences and the other from the SHBE domain \cite{shbe_domain}. Guided by researchers from these domains, we implemented workflows that integrated the researcher’s analytic scripts. The earth sciences use case included two workflows and the SHBE use case included three workflows.

Implementing a new use case in VAF includes three steps. First, we use the workflow/script management view to create the new use case in the use case management middleware repository. This step generates a unique use case key and associates it with a used-specified name. All subsequent workflows and their execution results will be associated with this key. The user also specifies the DAS data site(s) or hosts that will be leveraged by the workflows.

The second step involves reviewing the DAS configuration data. For VIFI, these data, stored in the \texttt{conf.yml} file and submitted to VIFI during task execution, specify the constraints that govern VIFI communications.

The third and final step for use case creation involves verifying that proper infrastructure constraints are satisfied. For example, proper firewall and security standards need verification with the organizations that will be hosting the VIFI infrastructure. Once a use case is created, workflows may be specified and executed, and results may be explored. In the following sections, we illustrate VAF through workflows from each evaluation use case. Figure \ref{fig:system-pipeline} denotes the technologies we leveraged for our implementation.

% this is a relatively simpler workflow. therefore, explaining this one first. then, move forward with the shbe use case
\subsection{Earth Science: Exploring Climate Projections}\label{sec:earth-science-use-case}

\begin{figure}[t]
    \centering
    \includegraphics[width=\linewidth]{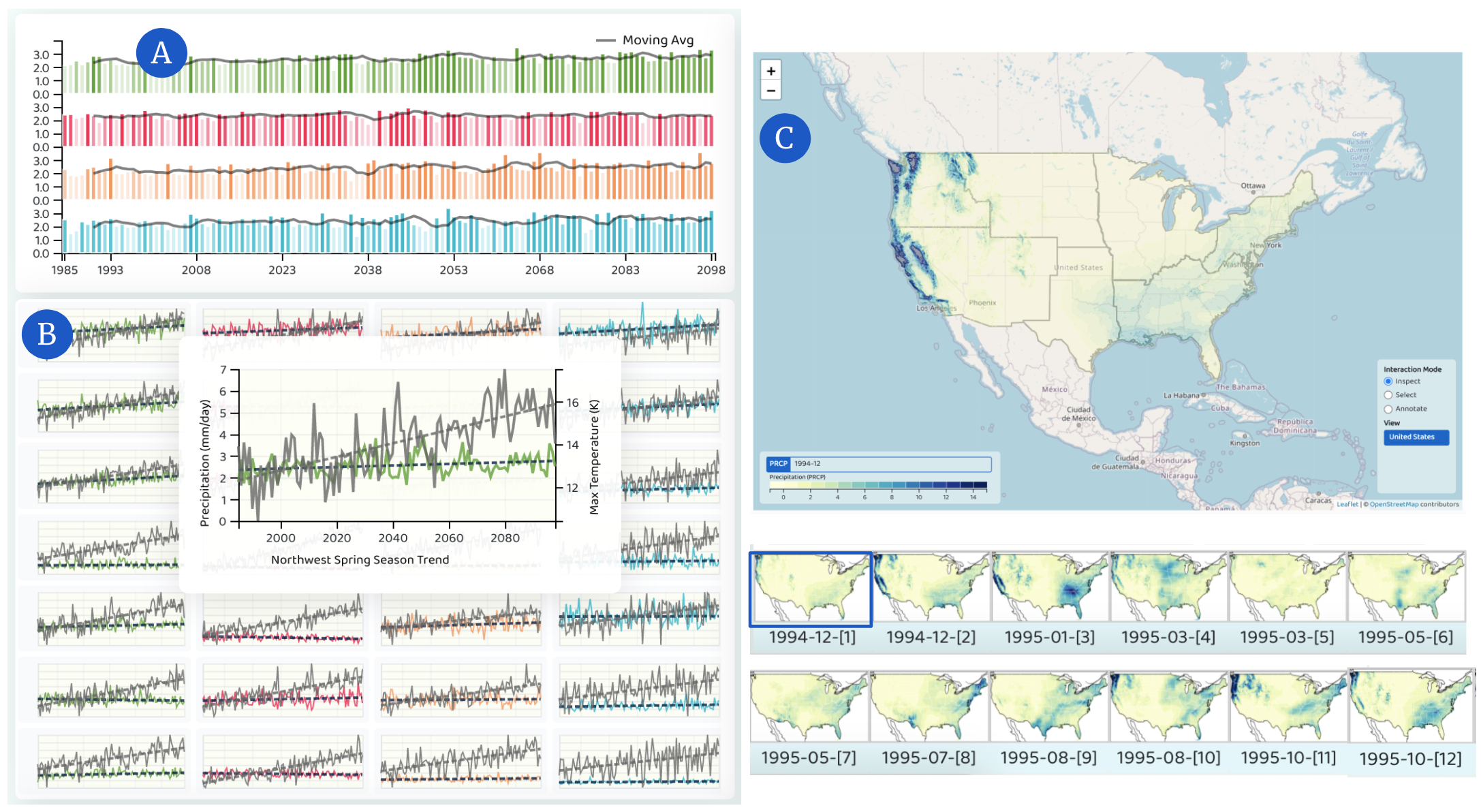}
    \caption{The visual analytic interface for the earth science use case, leveraging VIFI. Interactive geospatial visualization and trends for seasonal regional temperature and precipitation assist climate scientists in their analytic tasks.}
    \label{fig:earth-science-use-case}
\end{figure}

We used our VAF, leveraging the VIFI DAS, on NASA Earth Exchange published downscaled climate projections (NEX-DCP30) \cite{Wootten2021}. The United States National Climate Assessment (NCA) \cite{Jacobs2016} reports the future projections of the various climate variables from NEX-DCP30 to assess changing climate scenarios \cite{Nemani2015, Alder2015}. Recognizing its importance, the NASA Earth Exchange project released NEX-DCP30 data (observed and projected) that contain monthly averaged precipitation and temperature data for the contiguous US from 1985 to 2099. The projection data are stored in Network Common Data Form (NetCDF) \cite{edward2008experience} format and provide access to the projection output for 36 climate models \cite{Wootten2021}.

To perform demonstration evaluations of our VAF integrated with VIFI, we worked with a NASA climate scientist to develop workflows for analyzing NEX-DCP30. These workflows extracted the NetCDF data files and summarized monthly averaged spatiotemporal data for interactive, visual exploration. The first workflow executes data extraction analyses based on user provided parameters, such as projection model(s), climate variable(s), and year(s). An analytic script uses these parameters to find the corresponding NetCDF data and extracts geospatial contours for each month of the given year. The script and workflow configuration were authored and stored in the middleware using the Code Editor. The configuration file identifies the dataset (e.g., DEX-DCP30) and links via the middleware to authorization credential required for execution. In fact, the workflow configuration file contains all of the required parameters to execute this workflow. Hence, each time users execute a workflow, they update the parameters in the configuration file to extract the projection model of interest. The resulting data are formatted as GeoJSONs \cite{butler2016geojson}, subsequently stored in the middleware repository (e.g., an S3 bucket). Once data extraction is complete, the user can visualize and interactively explore the results as shown in Figure \ref{fig:earth-science-use-case}C. Recall that the VAF visualization library provides a generic map view that renders the geospatial contour visualizations. The geospatial navigator in Figure \ref{fig:earth-science-use-case}C is coordinated with the geospatial view, rendered using a configurable slider built-in the visualization library.

The second earth sciences workflow summarizes the spatiotemporal climate projections from NEX-DCP30 for exploration and analyses. This workflow contains multiple analytic scripts to summarize data from different perspectives while using different statistical techniques. Multiple scripts are included in this workflow since they share similar analysis goals. Users can reconfigure the workflow to use different scripts based on preference and interest. Workflow results contain monthly, seasonal, and yearly summaries of precipitation and temperature grouped by season and region. We created custom visualizations for this workflow as depicted in Figure \ref{fig:earth-science-use-case}. The requirement for this custom visualization was identified and co-designed by the participating climate scientist. Figure \ref{fig:earth-science-use-case}A shows multiple bar charts, sharing similar axes, illustrating the mean precipitation from 1985 to 2098, for each season. Figure \ref{fig:earth-science-use-case}B provides small multiples of precipitation and temperature trends for the 21st century. Each small multiple denotes a region and season correspondingly from top to bottom and left to right. 
% The custom visualization is designed from more primitive visualizations (e.g., simple line and bar charts). 
In this use case, the custom visualization can be used for exploration and analyses independent of the script that configures the workflow.

% describe workflow and goal of analytic scripts
% explain setup - data source, detail process
% how we triggered from the interface
% what was the resulted data
% how we visualized the result
%

\subsection{SHBE: Light Switching in Smart Buildings}\label{sec:use case 1: LSU_ANN}
\begin{figure}
    \centering
    \includegraphics[width=\linewidth]{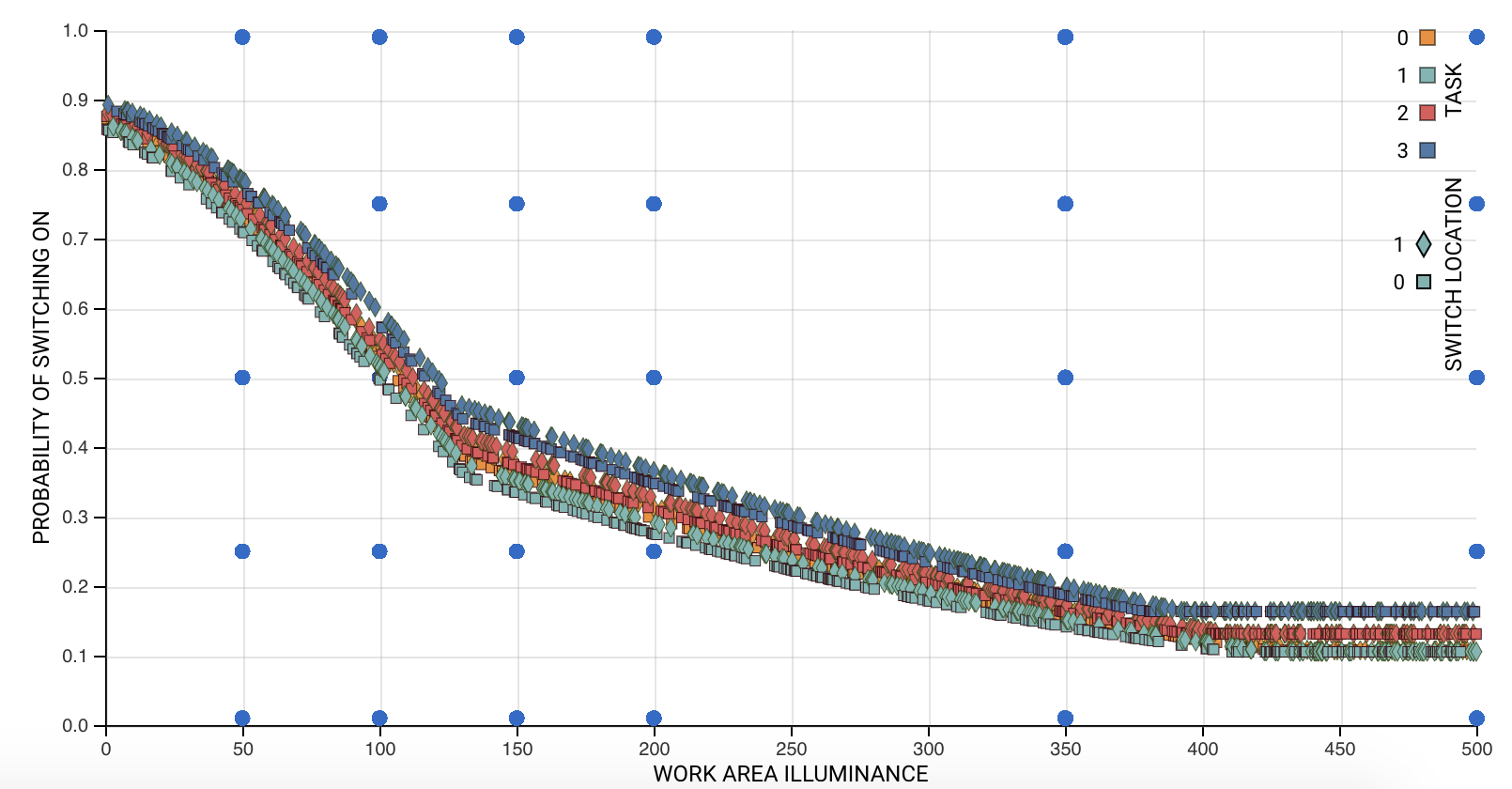}
    \caption{The SHBE use case result exploration for the smart building workflow. The scatterplot illustrates the light switching on-off probabilities based on the work area illuminance using the ANN model.}
    \label{fig:lsu-use-case-re}
\end{figure}

The SHBE domain is a multidisciplinary field that explores the interplay of human behaviors and the built environment with the goal toward a more sustainable future. Multiple workflows have been explored in collaboration with SHBE researchers. For space consideration, we highlight just one of these workflows to illustrate how more complex workflow designs are supported and enabled by VAF. The analytical purpose of the highlighted SHBE workflow is to explore the use and efficacy of Artificial Neural Networks (ANN) for the predication light on-off switching probabilities for the work area illuminance in a smart building as shown in the interactive VAF visualization presented in Figure \ref{fig:lsu-use-case-re}.
To illustrate the complexity of the analyses, we summarize the workflow implementation in VIFI below.

\begin{figure}
	\centering
	\includegraphics[width=\linewidth]{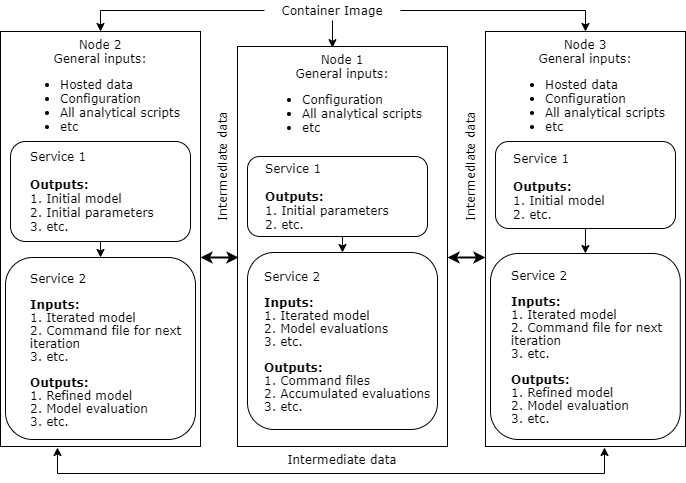}
	\caption{Workflow implementation for SHBE light switch on-off probability in smart buildings.} %Three workflows are developed to iteratively perform ANN for prediction.}
	\label{fig:lsu-use-case-implementation}
\end{figure}
%
% The VIFI implementation for this workflow is depicted in Figure \ref{fig:lsu-use-case-implementation}. 
As shown in Figure \ref{fig:lsu-use-case-implementation}, the workflow involves analysis over three distributed datasets at three different VIFI Data Sites. The data at each VIFI  Data Site is used by the ANN model for training and prediction. The third VIFI Data Site collects the updated ANN model and determines whether further model refinement is required using any of the other 2 VIFI Nodes. Thus, the third VIFI Node sends a different command file to each Node to specify what to do in the next step (e.g., fit the ANN model using existing data, use the ANN model to make predictions, etc.). Finally, when the third VIFI Node decides that the model is "good enough", the stopping condition is reached. The VIFI Orchestrator terminates the workflow and results are returned to the VAF middleware.

The ANN model, as well as other intermediate results, are sent between the VIFI Data Sites using RESTful API-based VIFI Orchestrator. The RESTful API is also used by each VIFI Data Site to accept incoming requests for analyses from  users launching workflow. Similar to the earth science use case, each request for analyses contains the required scripts, parameters, and workflow configuration. The configuration contains important information for proper workflow execution including the dataset(s), PAC(s), and input parameters as well as operation settings such as where to send intermediate and final results, whether to keep local copy of the (intermediate) results for further analysis, whether to add timestamps to results for potential time-series analyses, and other similar settings. In this workflow, analysis at each VIFI Data Site consists of two steps (or scripts). The output of each step is stored locally and transferred to other VIFI Data Sites for further processing. The first step in each Data Site in this workflow, executes only once but its output is used in multiple subsequent steps at this and other Data Sites. In other words, the initial ANN model is created at one of the VIFI Data sites as step one and it is used to predict outcomes and/or to train models at subsequent steps. Thus, it may execute any number of times until it is decided that the ANN model is “good enough” and the workflow is terminated. As mentioned previously, VAF supports the specification of the workflow and renders the output as a scatter plot as depicted in Figure \ref{fig:lsu-use-case-re}. This interactive visualization is customized so that square and diamond shaped glyphs denote switch-on and switch-off operations while color is used to denote independent workflow runs. The visualization describes the probability of light switch behavior for work area illuminance.

\section{Discussion and Limitation}
We presented a VAF for DAS to assist the data owners, researchers, and analysts to manage the infrastructure and conduct analysis through a web-based graphical UI. We have reviewed several distributed analysis systems such as XSEDE \cite{6866038}, SciServer \cite{medvedev2016sciserver}, and VIFI \cite{talukder2017vifi} to identify the design requirements to resolve the requirement for the user to directly access the server, manage the access control from the application layer, and facilitate the user to explore the result using interactive visualizations. 

We identified 7 implementation requirements that satisfies the design requirements to develop a web-based graphical UI for DAS. 
An interface for preparing the analytic scripts, configuring the workflow, and running the workflow in DAS sites resolves the requirement for the users to directly access the DAS servers. The middleware orchestrates the transactions between the UI and DAS. Moreover, the middleware manages the authentication and authorization from the application layer to reduce the workload of data owners. The workflows executed by the users through the UI are queued in the middleware database. The middleware communicates with the DAS sites to decide when to push the queued tasks and provide runtime and error logs to the UI that help the user to monitor the progress of the task. Finally, the visual exploration panel produces interactive visualizations to explore the resulted data from the analytic scripts.

We demonstrated the UI that satisfies the design requirements and illustrates the implementation requirements of our proposed VAF. The UI consists of 3 main panels - workflow management, task management, and visual exploration. The workflow management provides access to a hierarchical file structure (Figure \ref{fig:workflow-management}A), a component for creating or updating analytic scripts (Figure \ref{fig:workflow-management}B), and a terminal (Figure \ref{fig:workflow-management}C) to provide raw streaming logs. The middleware serves RESTful APIs to synchronize the UI with the DAS site on user's interactions. The task management panel provides status updates for the running workflows, overall runtime progress, and allows the user to either re-run the workflow or explore the result (Figure \ref{fig:task-management}). The visual exploration panel familiarizes the user with the data (Figure \ref{fig:result-exploration}), perceive their preferences to produce a set of interactive visualizations.

We implemented VAF in two use cases from earth science and SHBE domain. We leverage VIFI \cite{talukder2017vifi} DAS to implement 2 workflows from earth science and 3 workflows from SHBE. These workflows were initially configured and executed through a command line interface. The users were required to access multiple servers including the data sites to run their analyses. In contrast, after initial configuration and setup of VAF, the users are not required to access the distributed servers to create, update and run their analytical scripts. The pre-configured visual exploration panels for respected workflows assisted the analyst users to explore the result without any effort on creating interactive visualizations.

Nevertheless, we identified a few limitations of VAF based on our implementation experience. Our framework complies only with the DAS that provides RESTful APIs. 
% We have reviewed several popular distributed systems without any APIs or programmable APIs \cite{5360496, cohn2012dataone, pordes2007open} instead of modern RESTful APIs \cite{6866038, medvedev2016sciserver, talukder2017vifi} to be a candidate for our VAF. 
We plan to address this issue by developing a generic RESTful API and deploy at the DAS sites to comply with more distributed systems. The workflow configuration from the UI requires a learning curve for the users to be familiar with the configuration keywords
% and their roles in the workflow. 
We plan to provide a better interface with more readable labels and input validations for the configuration items which would ease the user with workflow configuration. 
We understand our visualization library lacks the use case specific visualization and interaction requirement to explore the results, which required workflow designers effort to preset the visualizations. We plan to create more input scopes for the users to inject their domain knowledge to influence the visualization recommendation \cite{vartak2017towards}.

\section{Conclusion}
In this paper, we presented a visual analytics framework (VAF) for distributed data analysis systems (DAS) to mediate user's direct interaction with the distributed servers, provide access control from application layer, and enable the exploratory visual analysis of results. To demonstrate the benefit of our proposed framework, we developed workflows for two use cases from earth science and SHBE research domains, working with respective domain experts. While we understand the potential of our VAF in distributed data analysis, we have several takeaways for future directions. Our future work will focus complying with more distributed systems developing a generic API service to deploy at DAS sites. Moreover, we aim to provide a more convenient interface for configuration management and perceive user's domain knowledge to provide interactive visualization recommendation to explore the resulted data. 

% \scriptsize{
\section*{Acknowledgment}
This paper was supported by the US National Science Foundations (NSF) Data Infrastructure Building Blocks
(DIBBs) Program (Award \#1640818).

\scriptsize{
\bibliographystyle{IEEEtran}
\bibliography{IEEEabrv,vifi_reference}
}
% \bibliography{vifi_reference}
\end{document}